\documentclass[aps,10pt,superscriptaddress,twocolumn,a4paper,prd]{revtex4-2}
\usepackage[colorlinks=true, pdfstartview=FitV, linkcolor=blue, citecolor=red, urlcolor=black]{hyperref}
\usepackage{graphicx}
\usepackage[all]{xy}
\usepackage{amsmath}
\usepackage{amssymb}
\usepackage{tensor}
\usepackage{orcidlink}
\newcommand{\be}{\begin{equation}}
\newcommand{\ee}{\end{equation}}
\newcommand{\ben}{\begin{eqnarray}}
\newcommand{\een}{\end{eqnarray}}
\newcommand{\bes}{\begin{subequations}}
\newcommand{\ees}{\end{subequations}}
\def\bal#1\eal{\begin{align}#1\end{align}}

\newcommand{\bfi}{\begin{figure}}
\newcommand{\efi}{\end{figure}}
\newcommand{\bc}{\begin{center}}
\newcommand{\ec}{\end{center}}

\newcommand{\sech}{\mbox{sech}}

\newcommand{\LL}{{\cal L}}

\begin{document}

\title{Scalar fields with impurities in arbitrary dimensions: \\first-order framework and exact solutions}
\author{D. Bazeia\,\orcidlink{0000-0003-1335-3705}}
     \email[]{bazeia@fisica.ufpb.br}
    \affiliation{Departamento de F\'\i sica, Universidade Federal da Para\'\i ba, 58051-970 Jo\~ao Pessoa, PB, Brazil}
    
        \author{M.A. Marques\,\orcidlink{0000-0001-7022-5502}}
        \email[]{marques@cbiotec.ufpb.br}
    \affiliation{Departamento de Biotecnologia, Universidade Federal da Para\'iba, 58051-900 Jo\~ao Pessoa, PB, Brazil}
    
    \author{R. Menezes\,\orcidlink{0000-0002-9586-4308}}
     \email[]{rmenezes@dcx.ufpb.br}
    \affiliation{Departamento de Ci\^encias Exatas, Universidade Federal da Para\'iba, 58297-000 Rio Tinto, PB, Brazil}
\affiliation{Departamento de F\'{\i}sica, Universidade Federal de Campina Grande,  58109-970 Campina Grande, PB, Brazil}

\begin{abstract}
We study a class of scalar field models coupled to impurities in arbitrary spacetime dimensions. The system admits the introduction of a second-order tensor that can be forced to obey an equality, if a first-order differential equation is satisfied, compatible with the equation of motion when the potential engenders a very specific form. In the case of static solutions, the energy density of the system can equal the divergence of an auxiliary vector function, which is included to help us solve the model. Stability of the field configuration under rescale of argument is investigated, and the procedure is illustrated considering distinct canonical models. The results show that exact solutions can be obtained in arbitrary dimensions, related to the presence of the first-order equation. 
\end{abstract}
\maketitle

In high energy physics, kinks, vortices and magnetic monopoles are localized structures of very general interest \cite{raja,manton}. Kinks are perhaps the simplest configurations, which appear in one spatial dimension in the presence of a single real scalar field with canonical kinematics and nonlinear self-interactions \cite{raja,manton,DB,vachaspati,Shnir}.
However, the study of kinks in $(D,1)$ spacetime dimensions has a caveat shown by Derrick's theorem \cite{derrick}; it unveiled that, after performing a rescale of argument in the scalar field solution, no stable localized configuration can be obtained in two or more spatial dimensions when the system is governed by canonical scalar field model.

Derrick's theorem is not an impassable obstacle. We can, for instance, consider the inclusion of the spatial coordinates in the Lagrangian density, as investigated in Ref.~\cite{prl}. There, the presence  of the radial coordinate in the potential of the scalar field has unveiled the presence of radially symmetric and stable solutions in arbitrary spatial dimensions. This result has motivated us to further investigate scalar field models when the coordinates are explicitly present in the action, and here we enlarge the scope of the study by considering the explicit dependence to be added from the presence of impurities in the system, as considered for more than 40 years now, in the specific case of kinks. Some of the first investigations has appeared in Refs.~\cite{imp1,imp2,imp3,imp4,imp5,imp6,imp7,imp8,imp9}, and the issue has gained further evidence in more recent studies, as presented in Refs. \cite{imp10,imp11,imp12,imp13,imp14,imp15,imp16,imp17,imp18}.
Another motivation to study scalar fields with impurities is related to the scattering of kinks, as suggested in Refs.  \cite{scattering1,scattering2,scattering3,scattering4}. An interesting effect, in particular, is the spectral wall phenomenon unveiled in \cite{scattering1}, where a well-localized region in space may trap or bounce back an incoming field configuration. 

In this Letter, we bring the novelty of developing a novel framework, which unveils the possibility of obtaining a first-order differential equation that solves the equation of motion in arbitrary dimensions. To achieve this result, we have to consider a specific model, in which the impurities are arranged in the form of a Minkowski vector $\sigma_\mu=\sigma_\mu(x)$ to enter the action of the scalar field $\phi(x)$ in addition to the standard vector $\partial_\mu\phi$ and also, with the help of another vector $G_\mu=G_\mu(\phi)$. Explicitly, the model is governed by the action
\be\label{ac1}
S = \int d^\nu x\,\LL(\phi,\partial_\mu\phi, \sigma_\mu),
\ee
where $\nu=1+D$, with $D$ standing for the number of spatial dimensions, and with the Lagrangian having the form
\be\label{lagrangian}
\LL =  \frac12\left(\partial_\mu\phi + \sigma_\mu\right)\left(\partial^\mu\phi + \sigma^\mu\right)+ \sigma_\mu G^\mu -V(\phi).
\ee
Extensions of this model to include $k$-field and Born-Infeld dynamics will be discussed elsewhere \cite{BMM}.
Here we see that the impurity vector describes the presence of impurities and contains explicit dependence on the spacetime coordinates, but it enters the model independently of the field configurations. The above system couples $\sigma_\mu$ and $G_\mu$ in a manner that each component of $G_\mu$ only exists if its corresponding $\sigma_\mu$ also exists. For instance, if one takes $\sigma_0=0$ it is required to have $G_0=0$ to keep the model consistent. In the above model, since
$ {\partial\LL}/{\partial(\partial^\mu\phi)}=\partial_\mu\phi + \sigma_\mu $, the equation of motion can be written as
\be\label{eomstd}
\partial_\mu\partial^\mu\phi+\partial_\mu \sigma^\mu = \sigma_\mu G^\mu_\phi - V_\phi,
\ee
where $V_\phi=dV/d\phi$ and $G^\mu_\phi=dG^\mu/d\phi$. One sees that, in the absence of impurities, the system turns into the case of the canonical model. 

The construction of this model is of current interest, and it suggests the introduction of the tensor ${\cal T}_{\mu\nu}$ with the form
\be\label{tensor}
\tensor{\cal T}{^\mu_\nu}=(\partial^\mu\phi + \sigma^\mu)\partial_\nu\phi-\delta^{\mu}_{\nu}{\LL},
\ee
which reduces to the standard energy-momentum tensor $T_{\mu\nu}$ when the impurity vector vanishes. It is not conserved, but we can make it obey the equality $\partial_\mu{\tensor{{\cal T}}{^\mu_\nu}}=0$ when the scalar field  solves the equation of motion and obeys the first-order equation
\be\label{foeq}
\partial_\mu\phi + \sigma_\mu + G_\mu=0,
\ee
which can be made compatible with the equation of motion if one further imposes the identity
\be\label{poten}
V(\phi) =-\frac12G_\mu G^\mu + C,
\ee
where $C$ is a constant of integration. Interestingly, the above framework has led us to construct the potential \eqref{poten} for \emph{time-dependent} fields and impurities compatible with the first-order equation \eqref{foeq}, together with $\partial_\mu\tensor{\cal T}{^\mu_\nu}=0$.

It is of interest to notice that, even though the tensor \eqref{tensor} correctly reproduces the usual energy-momentum tensor in the absence of impurities, we have to deal with it with due care. For instance, considering the component $\tensor{{\cal T}}{^{0}_{0}}$, one sees that it does not lead to the energy because of the presence of the time coordinate in the impurity. Instead, if one integrates it in the spatial coordinates, it gets the form of the Jacobi energy function \cite{mann}, which only leads to the energy if there is no explicit dependence on time. Remarkably, as we shall see below, the presence of the aforementioned tensor leads to novel results in the static case, which we now investigate.

As it is well established, in the study of localized structures one usually deals with time-independent quantities. In this sense, let us now focus attention on static fields and impurities. We then take $\sigma_0=0=G_0$, with $\vec{\sigma}=(\sigma^1,\sigma^2,\cdots,\sigma^D)$ and $\vec{G}=(G^1,G^2,\cdots, G^D)$.
The equation of motion and the first-order equation become
\be\label{eomlstatic}
\nabla\cdot\left(\nabla\phi-\vec{\sigma}\right) = V_\phi +\vec{\sigma}\cdot\vec{G}_\phi,
\ee
and
\be\label{foeqD}
\nabla\phi=\vec\sigma+ \vec{G}.
\ee
We remark that the above first-order equation is, actually, a set of $D$ partial differential equations in the form $\partial_i\phi = \sigma^i+G^i$, for $i=1,\ldots,D$.
Also, the components of $\tensor{\cal T}{^\mu_\nu}$ are now given by
\bes\label{emtcompstatic}
\begin{align}\label{t00static}
\tensor{{\cal T}}{^0_0}&=\frac12\left(\nabla\phi -\vec{\sigma}\right)\cdot\left(\nabla\phi -\vec{\sigma}\right)+V + \vec{\sigma}\cdot\vec{G},\\
\tensor{{\cal T}}{^0_i}&=\tensor{T}{^i_0}=0,\\ \label{tijstatic}
\tensor{{\cal T}}{^i_j}& = \left(\partial^i\phi + \sigma^i\right)\partial_j\phi \,\nonumber \\ 
&+ \delta^i_ j\left(\frac12\left(\nabla\phi -\vec{\sigma}\right)\cdot\left(\nabla\phi -\vec{\sigma}\right)+V + \vec{\sigma}\cdot\vec{G}\right).
\end{align}
\ees
Since we are dealing with time-independent quantities, we can write the energy integrating the energy density $\tensor{{\cal T}}{^0_0}$ in all space. This gives
\be\label{energy}
E=\int_{\mathbb{R}^D} d^Dx\, \tensor{{\cal T}}{^0_0},
\ee
where ${\mathbb{R}^D}$ represents the volume of the spatial portions of the spacetime under consideration. We can use the above expression to study Derrick's scaling argument and investigate stability under rescaling of the spatial coordinates by taking $\vec{x}\to{\color{red}\vec{y}=}\lambda\vec{x}$, which implies $\phi(\vec{x})\to\phi^{(\lambda)}(\vec{x})=\phi(\lambda\vec{x})$, with $\lambda$ being a real parameter. We denote the energy of the rescaled solution as $E^{(\lambda)}$, which is written below
\begin{equation}
\begin{aligned}
    E^{(\lambda)} &= \int_{\mathbb{R}^D} d^Dx\,\Bigg( \frac12\left(\nabla\phi^{(\lambda)} -\vec{\sigma}(\vec{x})\right)\cdot\left(\nabla\phi^{(\lambda)} -\vec{\sigma}(\vec{x})\right) \\
    &+V\big(\phi^{(\lambda)}\big) + \vec{\sigma}(\vec{x})\cdot\vec{G}(\phi^{(\lambda)})\Bigg).
\end{aligned}
\end{equation}
By defining the operator $\Tilde{\nabla}$ with the derivatives with respect to the components of $\vec{y}$, we can write
\begin{equation}
\begin{aligned}
    E^{(\lambda)}\! &=\! \int_{\mathbb{R}^D}\! \frac{d^Dy}{\lambda^{D}}\Bigg( \frac12\left(\lambda\Tilde{\nabla}\phi -\vec{\sigma}\bigg(\frac{\vec{y}}{\lambda}\bigg)\right)\cdot\left(\lambda\Tilde{\nabla}\phi -\vec{\sigma}\bigg(\frac{\vec{y}}{\lambda}\bigg)\right)\\
    &+V(\phi) + \vec{\sigma}\bigg(\frac{\vec{y}}{\lambda}\bigg)\cdot\vec{G}(\phi)\Bigg).
\end{aligned}
\end{equation}
Supposing that the minimum energy occurs for $\lambda=1$, we require that $\partial E^{(\lambda)}/\partial\lambda|_{\lambda=1}=0$, which imposes the constraint
\be\label{derrick1a}
\int_{\mathbb{R}^D} d^Dx\,\tensor{{\cal T}}{^i_{\;i}}=0,
\ee
where the index $i$ has to be summed over. If the solution satisfies this condition, it is stable under contractions and dilations, since $\partial^2 E^{(\lambda)}/\partial\lambda^2|_{\lambda=1}>0$ for the  model under consideration.

Let us now suppose that the energy density in Eq.~\eqref{t00static} can be written as a divergence, by taking $\tensor{{\cal T}}{^0_0}=\nabla\cdot {\vec W}$,  
where $\vec{W} = (W^1,W^2,\ldots,W^D)$ is an auxiliary function which depends only on $\phi$, such that the energy \eqref{energy} reads
\be\label{energyw}
E = \int_{\mathbb{R}^D}d^Dx\,\nabla\cdot \vec{W}.
\ee
Of course, $\vec{W}$ has to be chosen to lead to localized structures with finite energy. Now, using $\tensor{{\cal T}}{^0_0}=\nabla\cdot {\vec W}$ and the explicit form of $\tensor{{\cal T}}{^0_0}$, we get from Eq.~\eqref{tijstatic} that $\tensor{{\cal T}}{^i_j} = -G^i\partial_j\phi + \delta^i_j \,\nabla\cdot\vec{W}$, which gives 
\be\label{stress}
\tensor{{\cal T}}{^i_i} =(D\vec{W}_\phi-\vec{G})\cdot\nabla\phi,
\ee
where $\vec{W}_\phi=d{\vec W}/d\phi$. Therefore, we see that, to get localized $\tensor{T}{^i_i}$, both $\vec{G}$ and $\vec{W}_\phi$ must vanish asymptotically. We then suggest the direct connection
\be\label{gw}
\vec{G}=\vec{W}_\phi,
\ee
which leads to the potential
\be\label{vwcanonical}
V(\phi) = \frac12\,\vec{W}_\phi\cdot \vec{W}_\phi,
\ee
after taking to zero the constant $C$ that appears in Eq.~\eqref{poten}.
This choice of $\vec{G}$ leads to $\tensor{{\cal T}}{^i_i} = (D-1)\,\nabla\cdot\vec{W}$, which can be combined with the constraint \eqref{derrick1a} to impose the condition
\be\label{condenergy}
(D-1)\int_{\mathbb{R}^D}d^Dx\,\nabla\cdot\vec{W}=0.
\ee
This is required to get solutions which are stable against rescale of argument. Since the energy is given by Eq.~\eqref{energyw}, the above expression shows that the energy has to be zero for $D>1$. This shows that, in the model described by the Lagrangian \eqref{lagrangian}, the presence of the first-order equation and the impurity lead to a contribution that conspires against the energy of the field, making the total energy vanish. The case $D=1$ is special, and the energy may not be zero, being determined by the specific model under investigation. Interestingly, this fact is fully compatible with the results obtained before in Refs.~\cite{imp10,imp11,imp12,imp13}. Moreover, from Eq.~\eqref{stress}, one sees that $\tensor{{\cal T}}{^1_1}=0$ for any solution in $D=1$.

Let us now further consider the case of a \textit{single} spatial dimension, where both the impurity $\vec{\sigma}$ and $\vec{W}$ become scalars, so we denote them by $\sigma=\sigma(x)$ and $W=W(\phi)$. The equation of motion \eqref{eomlstatic} now takes the form 
\be\label{eomlstatic1D}
\left(\phi'-\sigma\right)' = V_\phi +\sigma W_{\phi\phi},
\ee
where the prime indicates derivative with respect to $x$. Also, the first-order equation \eqref{foeqD} becomes
\be\label{fostatic1D}
\phi^\prime=\sigma+ W_\phi,
\ee
and the potential gets to the form 
\be\label{poten1}
V=\frac12 W^2_\phi.
\ee
This leads us to the case of BPS states, as minimum energy configurations based on the procedure unveiled by Bogomol'nyi \cite{bogo} and by Prasad and Sommerfiled \cite{ps}.

 We notice, in particular, that the spatial derivative of the first-order equation \eqref{fostatic1D} leads us directly to the equation of motion \eqref{eomlstatic1D}, as expected. Moreover, the value of the energy depends on the model under investigation, and the expression in Eq.~\eqref{energyw} simplifies to
\be\label{ewD1}
E = W(\phi(\infty))-W(\phi(-\infty)),
\ee
which only depends on the function $W$ calculated at the asymptotic values of the solution of the first-order equation \eqref{fostatic1D}.

In higher dimensions, for $D>1$, the first-order equation \eqref{foeqD} reads
\be\label{focanonical}
\nabla\phi = \vec{\sigma} +\vec{W}_\phi.
\ee
It is compatible with the equation of motion \eqref{eomlstatic}
\be
\nabla^2\phi=\nabla\cdot\vec{\sigma} + \vec{W}_{\phi\phi}\cdot\vec{W}_{\phi}  +\vec{\sigma}\cdot\vec{W}_{\phi\phi},
\ee
if the potential has the form \eqref{vwcanonical}.
Also, we can write the energy density \eqref{t00static} as
\be\label{rhonablaphi}
\tensor{{\cal T}}{^0_0} = \left(\nabla\phi -\vec{\sigma}\right)\cdot\nabla\phi,
\ee
or, equivalently,
\be\label{rhophi}
\tensor{{\cal T}}{^0_0}= \left(\vec{W}_\phi + \vec{\sigma}\right)\cdot\vec{W}_\phi.    
\ee

We highlight here that this model also supports a BPS bound in the lines of Refs.~\cite{bogo,ps}. Indeed, one can write the energy density \eqref{t00static} in the form, using \eqref{gw} and \eqref{vwcanonical},
\be
\tensor{{\cal T}}{^0_0}=\frac12\left(\nabla\phi-\vec{\sigma}\right)^2 + \frac12\; \vec{W}_\phi\cdot\vec{W}_\phi + \vec{\sigma}\cdot\vec{W}_\phi,
\ee
or better
\be
\tensor{{\cal T}}{^0_0}
=\frac12\left(\nabla\phi-\vec{\sigma}-\vec{W}_\phi\right)^2 +\nabla\cdot\vec{W}.
\ee
Thus, if the first-order equation \eqref{focanonical} is satisfied, the energy density reduces to $\tensor{{\cal T}}{^0_0}=\nabla\cdot\vec{W}$ and the energy is minimized to the value given by Eq.~\eqref{energyw}. This is another way to confirm the stability of the solution.
t is also interesting to note that, since we are looking for spatially localized solutions, the first-order equation \eqref{focanonical} requires that we deal with the impurity with due attention.

We emphasize that the above procedure recovers the results obtained in Refs.~\cite{imp10,imp11,imp12,imp13} for static field in a single spatial dimension, with the very same first-order equation and potential. In higher dimensions, our model generalizes the recent result described in Ref.~\cite{imp17}, since here we can take the vector $\vec{W}=\vec{W}(\phi)$ with distinct components.
\begin{figure}[h!]
    \centering{
\includegraphics[width=0.75\linewidth]{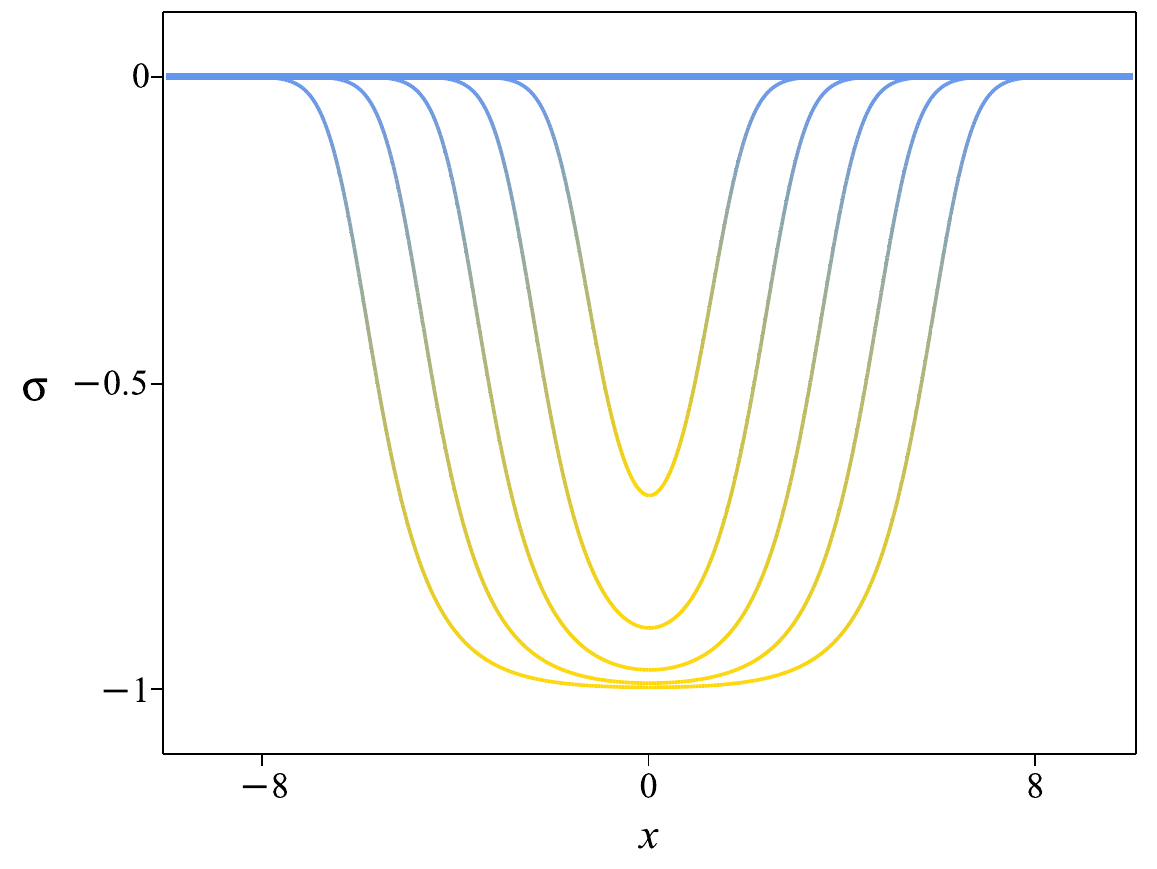}}
\caption{The impurity \eqref{sigma1} depicted for $a=0$ and for $a=0.9 ,0.99,0.999,0.9999,$ and $0.99999$. The thickest line represents the case $a=0$.}
\label{fig1}
\end{figure}

To illustrate the above results, let us first investigate a model in $D=1$. We consider spatially localized impurity in the form $\sigma(x)=f(x)\,\sech^2(x)$, with
\bal\label{sigma1}
&f(x)=\frac{\sqrt{1-a}-\sqrt{1-a\tanh^2(x)}}{\left(1-a\tanh^2(x)\right)^{3/2}}.
\eal
Here $a$ is a real parameter that obeys $a<1$. It is depicted in Fig.~\ref{fig1} and vanishes for $a=0$. One sees that as $a$ increases towards $1$, the impurity becomes a larger and larger well around the origin. 
\begin{figure}[t!]
    \centering{
\includegraphics[width=0.75\linewidth]{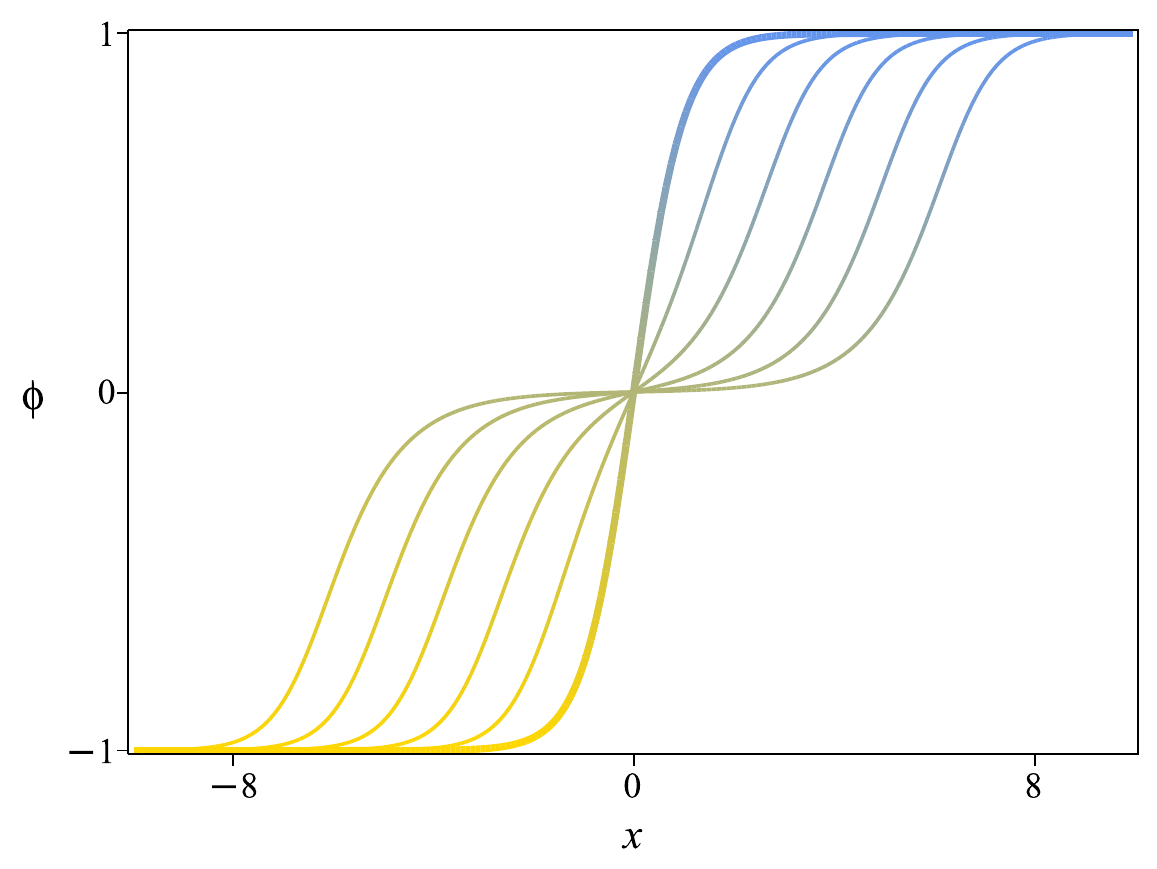}
\includegraphics[width=0.75\linewidth]{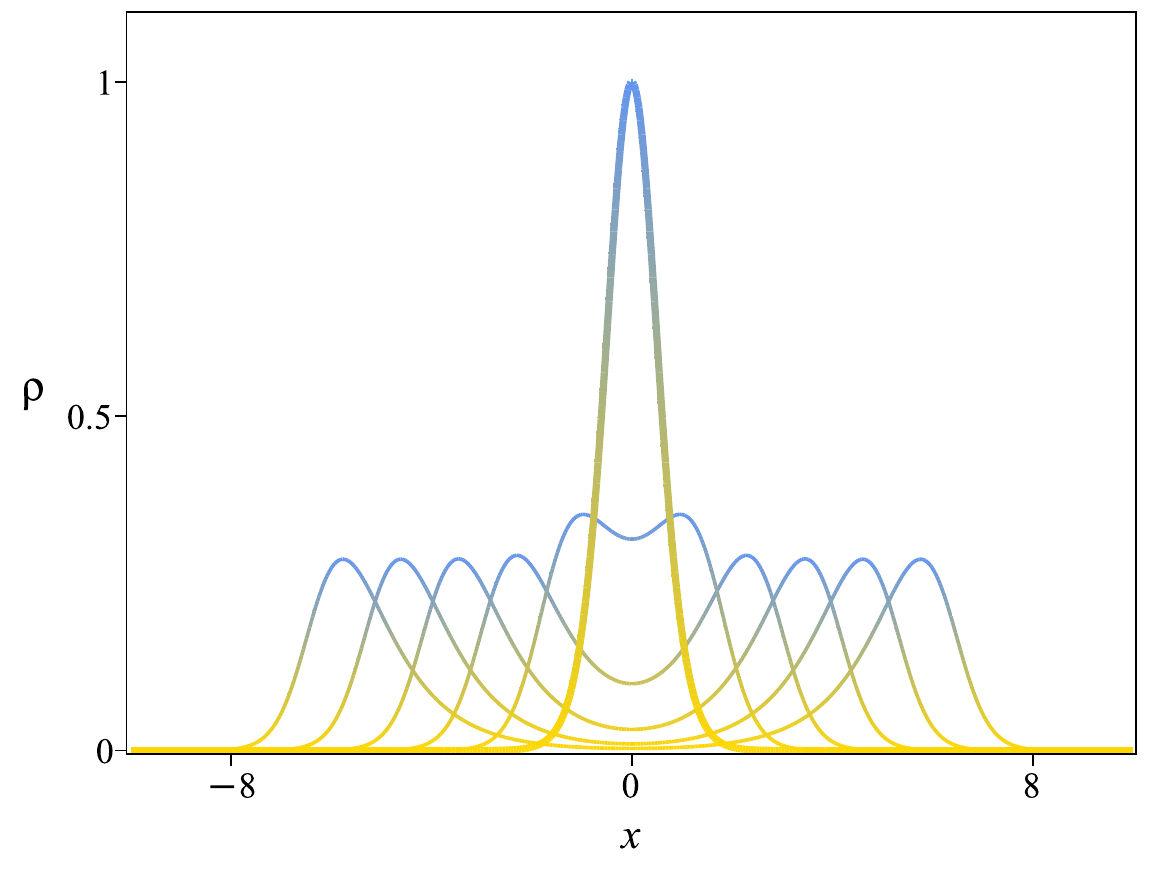}}
    \caption{The solution \eqref{solD1} (top) and its energy density \eqref{t00D1} (bottom), for the same values of $a$ considered in Fig. 1.}
    \label{fig2}
\end{figure}

We also take $W(\phi)=\phi-\phi^3/3$, which represents the $\phi^4$ model. We solve the first-order equation, seeking for solutions that asymptotically connect the uniform minima states $\bar\phi_\pm=\pm1$ of the potential. This gives
\be\label{solD1}
\phi(x) = \frac{\sqrt{1-a}\,\tanh(x)}{\sqrt{1-a\tanh^2(x)}},
\ee
where we have used the condition $\phi(0)=0$ to determine the constant of integration that arises in the process. This is an exact solution, which engenders the energy density
\be\label{t00D1}
\tensor{{\cal T}}{^0_0} = \frac{\sqrt{1-a}\,\sech^4(x)}{\left(1-a\tanh^2(x)\right)^{5/2}}.
\ee
After integrating the above expression, we get the energy $E=4/3$, which agrees with the value expected from Eq.~\eqref{ewD1}. We see that, although we are considering the $\phi^4$ potential, the impurity acts strongly, changing the solution and the energy density in an important manner. These results are displayed in Fig. \ref{fig2} and we notice that, as $a$ approaches $1$, the impurity affects the slope of the solution, which tends to get an inflection point with null derivative at its center, with the energy density unveiling an internal structure. Interestingly, solutions with profile similar to the one depicted in Fig. \ref{fig2} appeared before in Refs. \cite{prl,christ,BLM}. In the more recent work \cite{BLM}, it emerged under the action of a kinetic modification of the scalar field due to the presence of a second scalar field, which worked to generate a geometrical constriction,
as previously suggested in \cite{prb} in the study of domain walls in magnetic materials. Of course, here the modification of the internal structure of the kink is due entirely to the impurity, which acts as a well that strongly induces the scalar field to deviate importantly from the uniform states $\bar\phi_\pm=\pm1$ around the impurity location.

\begin{figure}[t!]
    \centering
\includegraphics[width=0.75\linewidth]{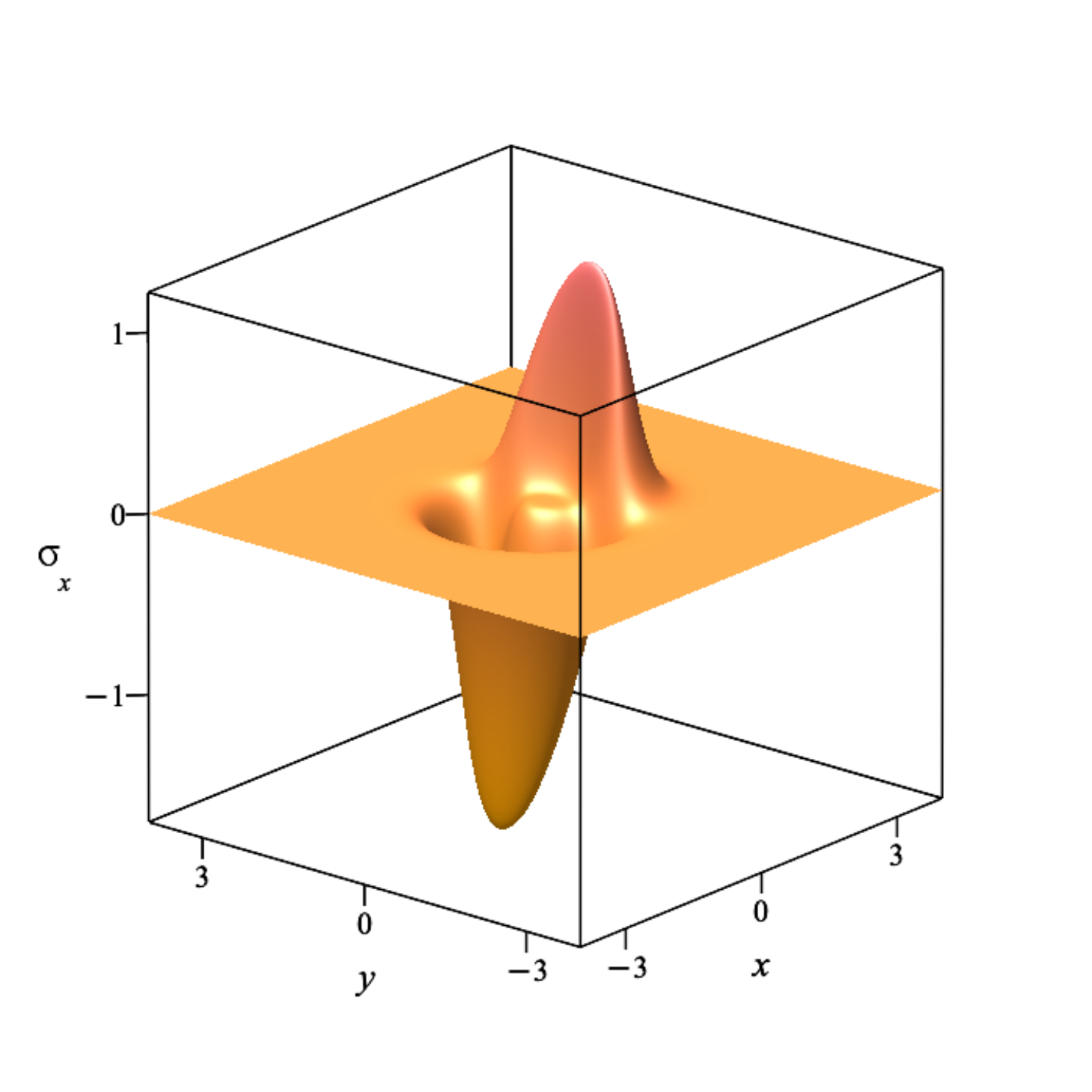}
    \caption{ The impurity $\sigma_x=\sigma^1(x,y)$ in \eqref{sigmax}.}
    \label{fig3}
\end{figure}

\begin{figure}[h!]
    \centering
\includegraphics[width=0.75\linewidth]{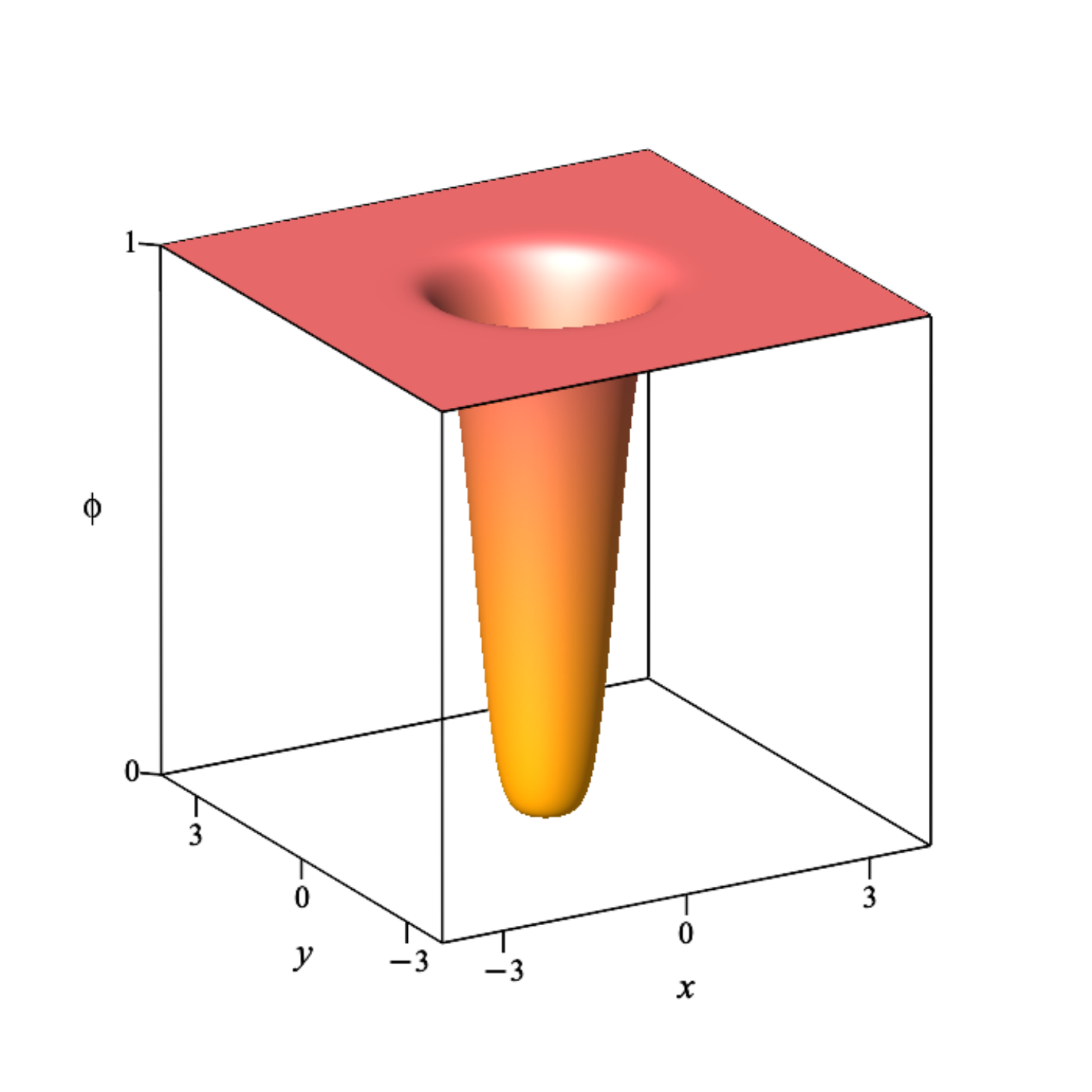}
\includegraphics[width=0.75\linewidth]{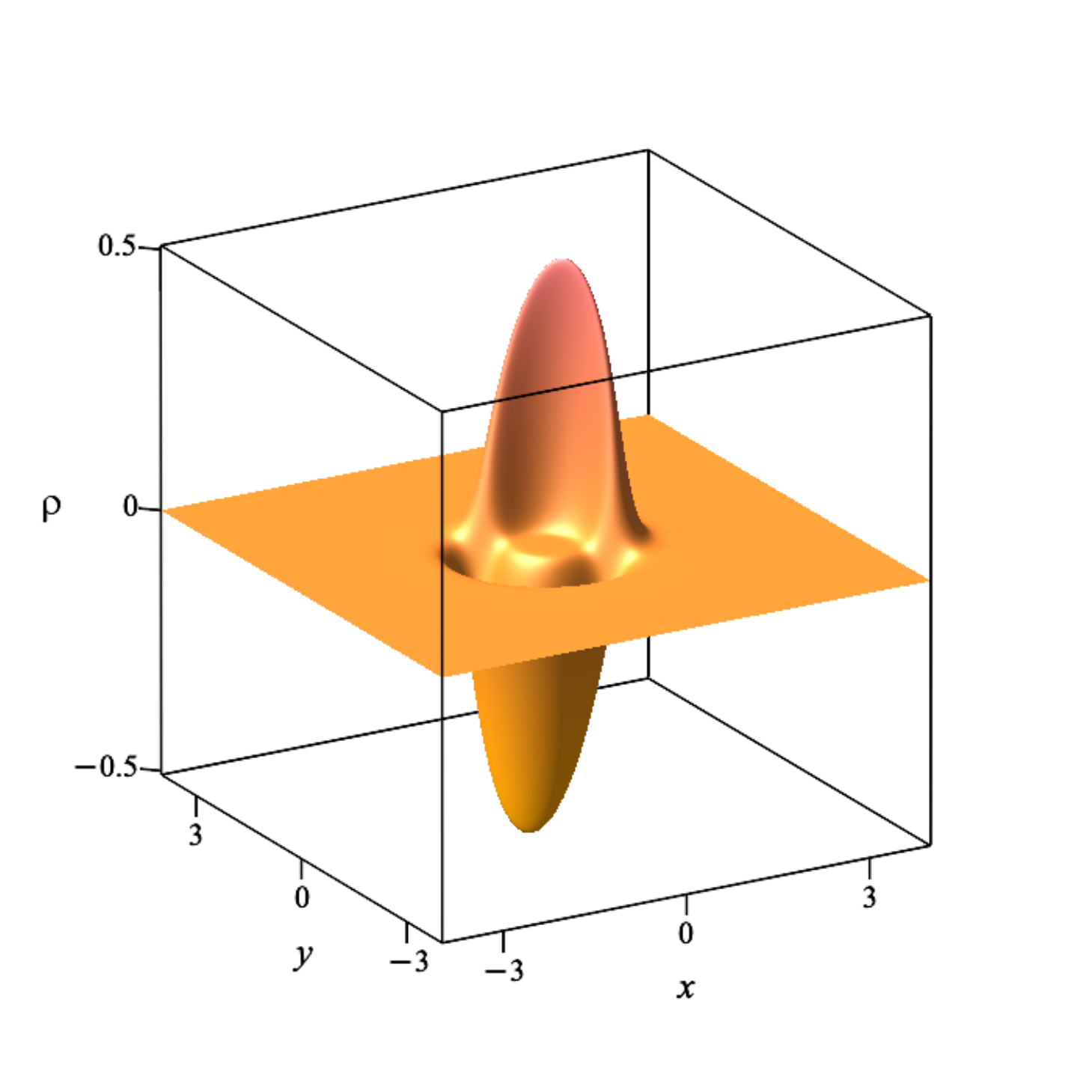}
    \caption{ The solution \eqref{solD2std} (top) and its energy density $\rho=\tensor{{\cal T}}{^0_0}(x,y)$ in \eqref{rhoD2std} (bottom).}
    \label{fig4}
\end{figure}

Let us now investigate localized structures in two spatial dimensions. In this situation, both the impurity and $\vec{W}$ have two components. We suppose that the impurity vector is localized around the origin, described by $\vec{\sigma}=(\sigma^1, \sigma^2)$, where
\bes
\begin{align}\label{sigmax}
\sigma^1(x,y)&=(6x - 1)\tanh^2(x^2\! + y^2)\,\sech^2(x^2\!+ y^2),\\
\sigma^2(x,y)&=(6y - 1)\tanh^2(x^2\! + y^2)\,\sech^2(x^2\! + y^2).
\end{align}
\ees
This choice allows for the presence of analytical results, leading to radially-symmetric field solutions, as we show below. In Fig.~\ref{fig3} we display $\sigma_x=\sigma^1(x,y)$ to illustrate how the impurity appears in this case. The other component $\sigma_y=\sigma^2(x,y)$ behaves similarly, changing $x\to y$ and $y\to x$. We also consider $\vec{W}=(W(\phi),W(\phi))$, with
\be\label{wpmodel}
W(\phi) = \frac{3}{5}\phi^{5/3} -\frac{3}{7}\phi^{7/3},
\ee
which leads to the potential
\be
V(\phi) = \phi^2\left( \phi^{1/3}-\phi^{-1/3}\right)^2.
\ee
This potential supports minima at $\phi=\pm1$ and $\phi=0$. It appeared before in \cite{prl}, and also in \cite{jcap}, where it was used to describe braneworld model having nontrivial internal structure. The first-order equation can be solved analytically; it engenders the exact solution
\be\label{solD2std}
\phi(x,y) = \tanh^3(x^2+y^2),
\ee
where we have taken the condition $\phi(0,0)=0$. The energy density can be written as
\be\label{rhoD2std}
\tensor{{\cal T}}{^0_0}(x,y) = 6(x+y)\tanh^4(x^2 + y^2)\,\sech^4(x^2 + y^2).
\ee
By integrating it, one finds that the energy is null, as expected. We display in Fig.~\ref{fig4} the solution \eqref{solD2std} and the energy density above. Interestingly, we notice that, even though the solution engenders radial symmetry, the energy density does not. This is because the impurity is not radially symmetric, strongly contributing to shape the energy density and nullify the total energy. We emphasize that despite the form of the impurity, the stability against rescale of the solution is ensured. However, we remark that other types of stability may be investigated. For instance, one may study the behavior of the solutions in the presence of small fluctuations, which is the so-called linear stability. The fact that the solution is stable against contractions and dilations does not ensure that it is also linearly stable. This issue will be thoroughly investigated in Ref.~\cite{BMM}, where we find solutions that are also linearly stable, including the solution \eqref{solD2std}.

In summary, we have investigated the specific class of canonical models \eqref{lagrangian}, described by Lagrangian that engenders a single real scalar field in the presence of impurities. Interestingly, the formalism have led us to a first-order equation which is compatible with the equation of motion if a constraint is satisfied. The formalism can be applied to both time-independent and time-dependent fields and impurities. In the case of time-independent quantities, it was possible  to introduce the vector function $\vec{W}(\phi)$ that allows us to write the energy as the integral of a divergence. By considering that $\vec{G}=\vec{W}_\phi$, we have shown that stable solutions require null energy for $D>1$, with the case $D=1$ being special, with the energy depending on the specific model under investigation. Also, we have shown that the model supports a BPS bound, and illustrated the method for specific sets of potential and impurity in one and two spatial dimensions, constructing the exact solutions explicitly.

There are others distinct lines of continuation of the present study. In particular, we are now extending the model to include generalized scalar fields, allowing for the presence of $k$-field and Born-Infeld dynamics, adding a complete study of stability of solutions of the first-order equation against rescale of argument, spatial translations and small fluctuations \cite{BMM}. Another possibility is to investigate time-dependent fields and impurities and the use of several fields. If one consider a time-dependent complex scalar field, it is possible to examine how the present study can be extended to the case of Q-balls. In the case of two scalar fields, we can consider one complex and the other real, to see how to generalize the Friedberg-Lee-Sirlin model \cite{FLS} to incorporate impurities in three spatial dimensions. If this is further connected with the study of fields and impurities on curved backgrounds \cite{Morris}, extensions including the Einstein-Friedberg-Lee-Sirlin model, which have been recently considered to investigate black holes, Q-balls, and boson stars \cite{blackhole,Qballs,bosonstars} indicate another line of current interest.\\

\acknowledgments{The authors would like to thank Matheus Liao for helpful discussions. This work is supported by the Brazilian agency Conselho Nacional de Desenvolvimento Cient\'ifico e Tecnol\'ogico (CNPq), grants Nos. 402830/2023-7 (DB, MAM and RM),  303469/2019-6 (DB), 306151/2022-7 (MAM) and 310994/2021-7 (RM).}


\end{document}